\documentclass[10pt]{emulateapj}
\usepackage{graphicx}
\usepackage{psfrag}
\newcommand\bld[1]{\mbox{\boldmath $#1$}}
\newcommand{\bnabla}{\bld{\nabla}}
\renewcommand{\bv}{\bld{v}}

\newcommand{\pdv}[2]{\frac{\partial#1}{\partial#2}}
\newcommand{\dv}[2]{\frac{d#1}{d#2}}

\newcommand{\be}{\begin{equation}}
\newcommand{\ee}{\end{equation}}
\newcommand{\bea}{\begin{eqnarray}}
\newcommand{\eea}{\end{eqnarray}}
\newcommand{\ct}{c_a}
\newcommand{\cto}{c_0}
\newcommand{\cs}{c_m}

\newcommand{\Pm}{P_m}

\newcommand{\z}{\alpha}
\newcommand{\zo}{\alpha_0}
\shortauthors{Johnson}
\shorttitle{}
\begin{document}

\title{Simple Waves in Ideal Radiation Hydrodynamics}

\author{Bryan M. Johnson}

\affil{Lawrence Livermore National Laboratory, L-023, 
7000 East Avenue, Livermore, CA 94550-9698}

\begin{abstract}
In the dynamic diffusion limit of radiation hydrodynamics, advection dominates diffusion; the latter primarily affects small scales and has negligible impact on the large scale flow. The radiation can thus be accurately regarded as an ideal fluid, i.e., radiative diffusion can be neglected along with other forms of dissipation. This viewpoint is applied here to an analysis of simple waves in an ideal radiating fluid. It is shown that much of the hydrodynamic analysis carries over by simply replacing the material sound speed, pressure and index with the values appropriate for a radiating fluid. A complete analysis is performed for a centered rarefaction wave, and expressions are provided for the Riemann invariants and characteristic curves of the one-dimensional system of equations. The analytical solution is checked for consistency against a finite difference numerical integration, and the validity of neglecting the diffusion operator is demonstrated. An interesting physical result is that for a material component with a large number of internal degrees of freedom and an internal energy greater than that of the radiation, the sound speed increases as the fluid is rarefied. These solutions are an excellent test for radiation hydrodynamic codes operating in the dynamic diffusion regime. The general approach may be useful in the development of Godunov numerical schemes for radiation hydrodynamics.
\end{abstract}


\section{Introduction}

The equations of radiation hydrodynamics (RHD) are used to model fluids in which matter and radiation are coupled. In an opaque material, the coupling is strong and acts primarily to maintain the material and radiation fluids in local thermodynamic equilibrium (LTE). Once at a common temperature, the radiation acts as an additional source of pressure as well as a diffusive energy sink, and the material advects the radiation. The relative importance of diffusion and advection depends upon both the optical depth of the system and its characteristic velocity. If $\beta \tau \ll 1$, where $v = \beta c$ is the fluid velocity and ${\it l} = \tau \lambda_p$ is a characteristic length ($\lambda_p$ being a photon mean-free path), diffusion dominates advection; such a system is referred to as being in the static diffusion limit. If $\beta \tau \gg 1$, advection dominates diffusion and the system is in the dynamic diffusion limit. This paper is concerned primarily with dynamic diffusion.

In the dynamic diffusion limit, diffusion only operates on length scales $\lambda \ll {\it l}$, in a manner analogous to viscosity and heat conduction (albeit on different scales). The bulk flow properties can thus be modeled accurately by treating the radiation fluid as ideal, i.e., by neglecting radiative diffusion along with other forms of dissipation. Many of the methods developed for analyzing fluids without radiation can thus be applied to a radiating fluid operating in the dynamic diffusion limit. The effects of diffusion are limited to boundary layers and shocks and other regions of the flow in which gradients are large.

This viewpoint is pursued here by analyzing the equations of ideal RHD in one dimension. I begin in \S\ref{EQ} by outlining the equations of ideal RHD in the gray diffusion limit in a frame comoving with the fluid. A review of the thermodynamics of a radiating fluid is provided in \S\ref{THERMO}, and the impact of radiation on a centered rarefaction wave is presented in \S\ref{CRW}. Results from a finite difference numerical integration are included as a consistency check, along with an example of the effects of dissipation. A discussion of how to generalize this approach to include more general boundary conditions is given in \S\ref{CRI}, along with expressions for the characteristic curves and Riemann invariants of the combined fluid. A summary and a discussion of how these ideas might be used in the development of Godunov schemes for numerical RHD is given in \S\ref{SD}.

\section{Equations}\label{EQ}

The energy equations for an optically-thick gray medium in LTE, in a frame comoving with the fluid and neglecting scattering, are
\be\label{EM}
\dv{U}{t} - \gamma \frac{U}{\rho}\dv{\rho}{t} = c\kappa\left(E - aT^4\right)
\ee
and
\be\label{ER}
\dv{E}{t} - \frac{4}{3} \frac{E}{\rho}\dv{\rho}{t} = c\kappa\left(aT^4 - E\right) + \frac{c}{3}\bnabla \cdot \left(\frac{1}{\kappa}\bnabla E\right),
\ee
where $\rho$, $U$ and $T$ are the mass density, internal energy density and temperature of the material, $E$ is the energy density of the radiation, $c$ is the speed of light, $a$ is the radiation constant, $\kappa \sim \lambda_p^{-1}$ is the absorption opacity in units of inverse length, and $d/dt$ is the derivative following a fluid element. Equation (\ref{EM}) is specific to an ideal gas equation of state with material pressure
\be\label{EQSM}
\Pm = (\gamma - 1)U,
\ee
where $\gamma$ is the adiabatic index of the material.

The three relevant time scales for these equations are the advection time scale $t_{adv} = {\it l}/v$, the diffusion time scale $t_{diff} = \kappa {\it l}^2/c$ and the time scale for coupling between the material and radiation $t_{coup} = (c\kappa)^{-1}$ (the mean time between photon collisions). Their relative scalings are
\be
\frac{t_{adv}}{t_{coup}} \sim \frac{\tau}{\beta},
\ee
\be
\frac{t_{diff}}{t_{coup}} \sim \tau^2
\ee
and
\be
\frac{t_{diff}}{t_{adv}} \sim \beta \tau,
\ee
so that in the dynamic diffusion limit ($\tau \gg \beta^{-1}$)
\be
t_{coup} \ll t_{adv} \ll t_{diff}.
\ee

Thermal equilibrium is thus established on a short time scale, and diffusion can be neglected for spatial variations $\lambda \sim {\it l}$. This can be seen explicitly by subtracting equation (\ref{ER}) from equation (\ref{EM}):
\bea\label{SUB}
E - aT^4 = \frac{1}{2c\kappa}\left(\dv{}{t}\left[U - E\right] - \left[\gamma U - \frac{4}{3} E \right]\frac{1}{\rho}\dv{\rho}{t} \right. \nonumber \\ \left. + \frac{c}{3}\bnabla \cdot \left[\frac{1}{\kappa}\bnabla E\right]\right).
\eea
The diffusion operator is $O([\beta \tau]^{-1})$ relative to the advection terms, which are in turn $O(\beta \tau^{-1})$ relative to the left hand side, so that
\be\label{TB}
E \simeq aT^4
\ee
in the dynamic diffusion limit. Expression (\ref{TB}) will generally be considered to be an equality throughout this analysis, although of course departures from thermal balance, no matter how small, are necessary to maintain coupling between the radiation and the material. Denoting the total pressure $\Pm + P_r$ by $P$, where
\be\label{EQSR}
P_r = \frac{E}{3} \simeq \frac{aT^4}{3},
\ee
the equations of ideal RHD are given by
\be\label{MASS}
\frac{1}{\rho}\dv{\rho}{t} + \bnabla \cdot \bv = 0,
\ee
\be\label{MOM}
\rho\dv{\bv}{t} + \bnabla P = 0,
\ee
\be\label{EMI}
\dv{U}{t} - \gamma \frac{U}{\rho}\dv{\rho}{t} = c\kappa\left(E - aT^4\right),
\ee
and
\be\label{ERI}
\dv{E}{t} - \frac{4}{3} \frac{E}{\rho}\dv{\rho}{t} = c\kappa\left(aT^4 - E\right).
\ee
Coupled with the equations of state (\ref{EQSM}) and (\ref{EQSR}) for the material and radiation, these equations form a closed set. Their form in one Cartesian dimension is the basis for the analysis in the following sections.

\section{Thermodynamics}\label{THERMO}

Most of the contents of this section are not new; the results can be found in standard textbooks \citep{ch67,cg68,mm84} and are only included here for completeness and ease of reference, and to highlight the most general features of the solutions to be described in the next section. The sum of equations (\ref{EMI}) and (\ref{ERI}) is the first law of thermodynamics for the combined material-radiation fluid:
\be
d(U + E) - \left(\gamma U + \frac{4}{3} E\right)\frac{d\rho}{\rho} = 0.
\ee
It is straightforward to demonstrate that this is equivalent to
\be
d\left(s_m + s_r\right) = 0,
\ee
where
\be
s_m = \frac{k_B}{\mu m_p} \ln\left(\frac{U}{\rho^\gamma}\right)^\frac{1}{\gamma - 1}
\ee
is (to within a constant) the specific entropy of the material and
\be
s_r = \frac{4aT^3}{3\rho}
\ee
is the specific entropy of the radiation (i.e, the entropy of radiation per unit mass of material).\footnote{Notice that $s_r$ is the entropy of a photon gas coupled to a material fluid; it is not in general equivalent to $s_m$ with $\gamma = 4/3$. The two are equivalent (to within a constant factor) only in the high energy density limit.} Changes in ideal RHD are adiabatic: the total entropy of the combined fluid is conserved for a fluid element.

The first law combined with the equation of state for the material ($U \sim \rho T$) implies
\be\label{TRHO}
\left(\frac{1}{\gamma - 1} + 12\z\right)\frac{dT}{T} = \left(1 + 4\z\right) \frac{d\rho}{\rho},
\ee
where
\be\label{ZDEF}
\z \equiv \frac{P_r}{\Pm} \propto \frac{T^3}{\rho}
\ee
is the ratio of radiation to material pressure. The density can be expressed as a function of $\z$ by combining (\ref{TRHO}) and (\ref{ZDEF}): 
\be\label{DRDZ}
\frac{d\z}{\z} = 3\frac{dT}{T} - \frac{d\rho}{\rho} = \frac{3\gamma - 4}{1 + 12(\gamma - 1)\z}\frac{d\rho}{\rho},
\ee
or
\be\label{RHOZ}
\left(\frac{\rho}{\rho_0}\right)^{3\gamma - 4} = \frac{\z}{\zo}  e^{12(\gamma - 1)\left(\z - \zo\right)},
\ee
where a subscript denotes a reference value; for the problem analyzed in the following section, it is the quiescent value before the rarefaction wave is excited. An important implication of the above expression is that rarefaction (compression) is associated with an increase (decrease) in the pressure ratio $\z$ for $\gamma < 4/3$ (i.e., $\rho \lessgtr \rho_0$ when $\z \gtrless \zo$); for $\gamma > 4/3$, the opposite conditions hold. 

The other fluid quantities are given by
\be
\left(\frac{T}{T_0}\right)^{3\gamma - 4} = \left(\frac{\z}{\zo}\right)^{\gamma - 1} e^{4(\gamma - 1)\left(\z - \zo\right)},
\ee
\be
\left(\frac{\Pm}{P_0}\right)^{3\gamma - 4} = \left(\frac{1}{1 + \zo}\right)^{3\gamma - 4}  \left(\frac{\z}{\zo}\right)^\gamma e^{16(\gamma - 1)\left(\z - \zo\right)}
\ee
and
\be\label{PRZ}
\left(\frac{P_r}{P_0}\right)^{3\gamma - 4} = \left(\frac{\zo}{1 + \zo}\right)^{3\gamma - 4} \left(\frac{\z}{\zo}\right)^{4(\gamma - 1)} e^{16(\gamma - 1)\left(\z - \zo\right)}.
\ee
Expressions (\ref{RHOZ})-(\ref{PRZ}) match expressions $9.127$ of \cite{cg68} for $\gamma = 5/3$ and $\z = Z$.

For $\gamma = 4/3$, $\z = const.$ and expressions (\ref{RHOZ})-(\ref{PRZ}) must be replaced with
\be
\frac{\rho}{\rho_0} = \left(\frac{T}{T_0}\right)^3,
\ee
\be
\frac{\Pm}{P_0} = \frac{1}{1+\zo}\left(\frac{T}{T_0}\right)^4
\ee
and
\be
\frac{P_r}{P_0} = \frac{\zo}{1+\zo}\left(\frac{T}{T_0}\right)^4.
\ee

Small changes in the total pressure satisfy the following expression:
\be
dP = \Pm\left(\frac{d\rho}{\rho} + \left[1 + 4\z\right]\frac{dT}{T}\right).
\ee
Combining the above expression with expression (\ref{TRHO}) gives the adiabatic sound speed for the radiating fluid:
\be\label{CTDEF}
\ct^2 \equiv \left(\pdv{P}{\rho}\right)_{s} = \Gamma_1 \frac{P}{\rho},
\ee
where the total entropy is held constant and
\be\label{GDEF}
\Gamma_1 \equiv \dv{\ln P}{\ln \rho} = \frac{1}{1+\z}\left(1 + \frac{[\gamma - 1][1 + 4\z]^2}{1 + 12[\gamma - 1]\z}\right).
\ee
The closed form expression for $\ct$ as a function of $\z$ is
\be\label{CT}
\ct = \cto\left(\frac{\z}{\zo}\right)^{\frac{\gamma - 1}{2(3\gamma - 4)}} e^{\frac{2(\gamma - 1)}{3\gamma - 4}\left(\z - \zo\right)}\sqrt{\frac{(1+\z)\Gamma_1}{(1+\zo)\Gamma_0}},
\ee
where $\Gamma_0$ is $\Gamma_1$ evaluated at $\z = \zo$. The above expression can be combined with expression (\ref{ZRHO}) to obtain a closed form expression for $P(\rho)$; one can thus regard the radiation in ideal RHD as a modification of the equation of state for the fluid.

\section{Centered Rarefaction Wave}\label{CRW}

This section will attempt to follow closely the standard hydrodynamic analysis, a cogent expression of which can be found in \S\S99-105 of \cite{ll87}. One of the simplest one-dimensional flows to analyze is a centered rarefaction wave. The prototypical example of such a flow is that generated behind a piston moving out of a semi-infinite fluid at constant velocity. Since there are no characteristic time or length scales in such a system, all of the flow variables must depend upon the coordinates through the similarity variable $\xi = x/t$, so that $\partial/\partial t = -(\xi/t)d/d\xi$ and $\partial/\partial x = (1/t) d/d\xi$. The form of equations (\ref{MASS}) and (\ref{MOM}) for a one-dimensional similarity flow is
\be\label{S1}
(v - \xi)\dv{\rho}{\xi} + \rho \dv{v}{\xi} = 0,
\ee
\be\label{S2}
(v - \xi) \dv{v}{\xi} + \frac{1}{\rho}\dv{P}{\xi} = 0.
\ee

Combining these equations with equation (\ref{CTDEF}) gives
\be\label{CHAR}
v \pm \ct = \xi 
\ee
and
\be\label{V}
v = \pm \int \frac{\ct \, d\rho}{\rho} = \pm \int \frac{dP}{\rho \ct},
\ee
where a rarefaction wave corresponds to the plus sign and a compression wave to the minus sign. These correspond to the hydrodynamic results with $\cs = \sqrt{\gamma \Pm/\rho} \rightarrow \ct$ and $\Pm \rightarrow P$. Details on evaluating the above integral can be found in the Appendix; an approximate solution valid for $\gamma \gtrsim 1.1$ can be obtained from the following considerations.

From (\ref{CTDEF}) and (\ref{V}), 
\be
\frac{2}{\Gamma_1 - 1}\frac{d\ct}{\ct} = \frac{dv}{\ct} + \frac{d\Gamma_1}{\Gamma_1(\Gamma_1 - 1)}.
\ee
The above expression is exact. Since $\Gamma_1$ varies much more slowly with $\xi$ than either $v$ or $\ct$ ($\gamma < \Gamma_1 < 4/3$ and from [\ref{CHAR}] one expects $v$ and $\ct$ to vary approximately linearly with $\xi$), it can be approximated by
\be
\dv{v}{\ct} \simeq \frac{2}{\Gamma_0 - 1},
\ee
so that
\be\label{VAPPROX}
v \simeq \frac{2}{\Gamma_0 - 1}\left(\ct - \cto\right) \simeq \frac{2}{\Gamma_0 + 1} \left(\frac{x}{t} - \cto\right) 
\ee
and
\be\label{CAPPROX}
\ct \simeq \frac{2}{\Gamma_0 + 1}\cto + \frac{\Gamma_0 - 1}{\Gamma_0 + 1} \frac{x}{t},
\ee
where the velocity is defined to be zero in the initial state. This corresponds to the hydrodynamic result with $\cs \rightarrow \ct$ and $\gamma \rightarrow \Gamma_0$. The velocity slope varies between
\be
\frac{2}{\gamma + 1} < \dv{v}{\xi} < \frac{6}{7}.
\ee
To obtain the solution for the other flow variables, it is necessary to solve expressions (\ref{CT}) and (\ref{CAPPROX}) implicitly to obtain $\z(\xi)$, and then insert the result into expressions (\ref{RHOZ})-(\ref{PRZ}). 

The wave front is located at $x = \cto t$ and propagates away from the piston. A region of constant velocity equal to the velocity of the piston is located between the piston and the point
\be
x_{tr} = \left(\cto - \frac{\Gamma_0 + 1}{2} |v_p|\right) t,
\ee
where $v_p < 0$ is the piston velocity. The location of this transition between the similarity flow and the region adjacent to the piston travels away from the piston for $|v_p| < 2\cto/(\Gamma_0 + 1)$; otherwise it travels towards the piston. For a piston velocity greater in magnitude than
\be\label{VCRIT}
|v_p|_{crit} = \frac{2}{\Gamma_0 - 1} \, \cto,
\ee
the fluid is evacuated between the piston and the point $x = -|v_p|_{crit} t$ (the fluid quantities go to zero).

\begin{figure}
\psfrag{y}[][][1.5]{$\dv{v}{\xi}$}
\psfrag{x}[][][1.2]{$\log_{10} \zo$}
\plotone{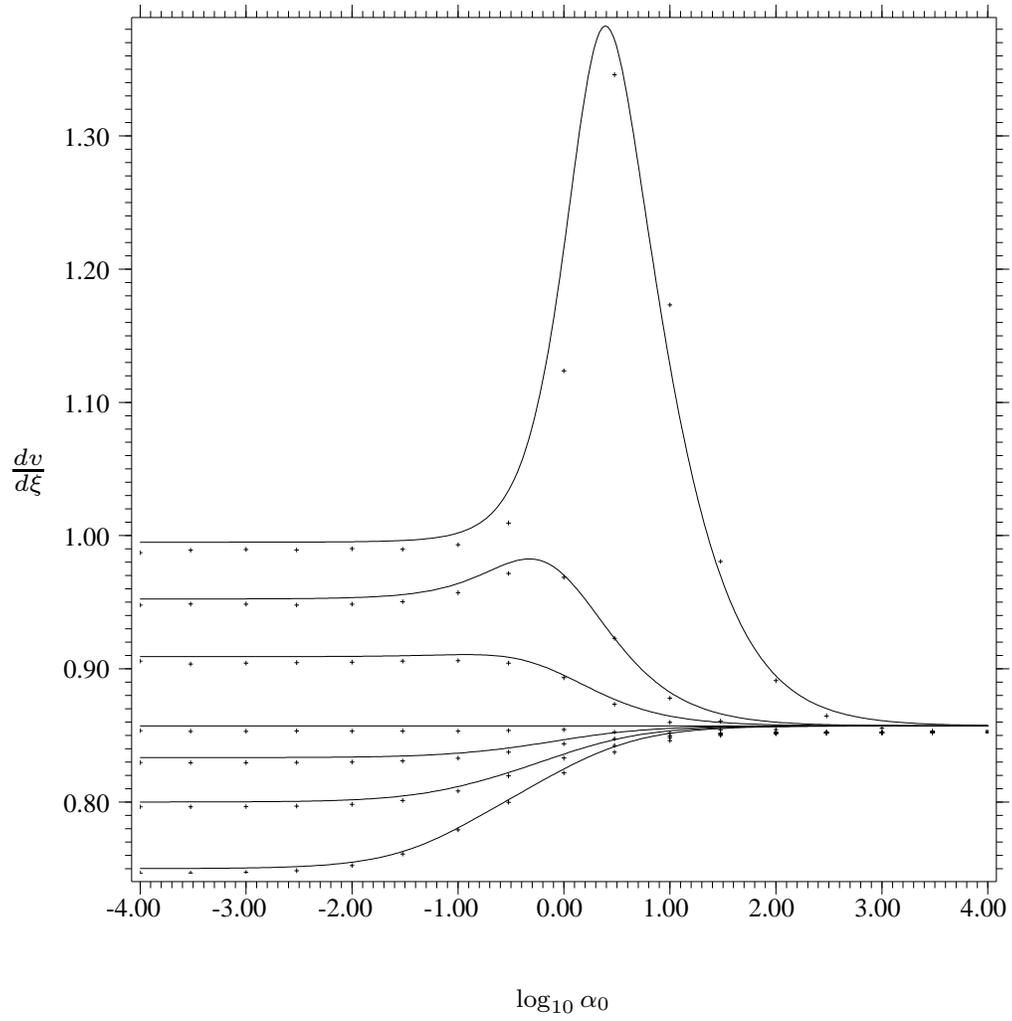}
\caption{
Velocity slope for $\gamma = 1.6667$, $1.5$, $1.4$, $1.3333$, $1.2$, $1.1$ and $1.01$ (from bottom to top). The solid lines are the approximate analytical results and the points are results from a finite difference numerical integration with $512$ grid points.
}
\label{slope}
\end{figure}

Figure~\ref{slope} shows the slope of the velocity as a function of $\zo$, the quiescent ratio of radiation to material pressure, for various values of $\gamma$.\footnote{Comparison to the exact solution indicates that expression (\ref{VAPPROX}) is accurate to within $0.8\%$ for $\gamma \geq 1.1$.} Also plotted in Figure~\ref{slope} are points from a finite difference numerical integration of equations (\ref{MASS})-(\ref{ERI}). The code used to obtain these results is a one-dimensional version of the ZEUS algorithm \citep{sn92,smn92,ts01} without diffusion. The initial temperature was set to give $\beta_0 \equiv \cto/c = 10^{-4}$ (this along with the value for $\zo$ sets the initial density), and the piston velocity was set to $\cto/(\Gamma_0 + 1)$. Power-law fits to the Rosseland mean opacity were used \citep{bl94}, and the computational domain was set to $L = \kappa_0^{-1}$, where $\kappa_0(\rho_0,T_0)$ is the initial opacity.\footnote{This corresponds to an optical depth of unity across the computational domain. Without diffusion, the only constraint on the overall length scale is $\tau_0 \equiv L \kappa_0 \gg \beta_0$ so that the coupling time scale is much less than the advection time scale.} The slope was measured from a least squares fit to the velocity profile as a function of $x$ after the wave front had propagated across the computational domain.

\begin{figure}
\psfrag{y}[][][1.5]{$\frac{v}{\cto}$}
\psfrag{x}[][][1.5]{$\frac{\xi}{\cto}$}
\plotone{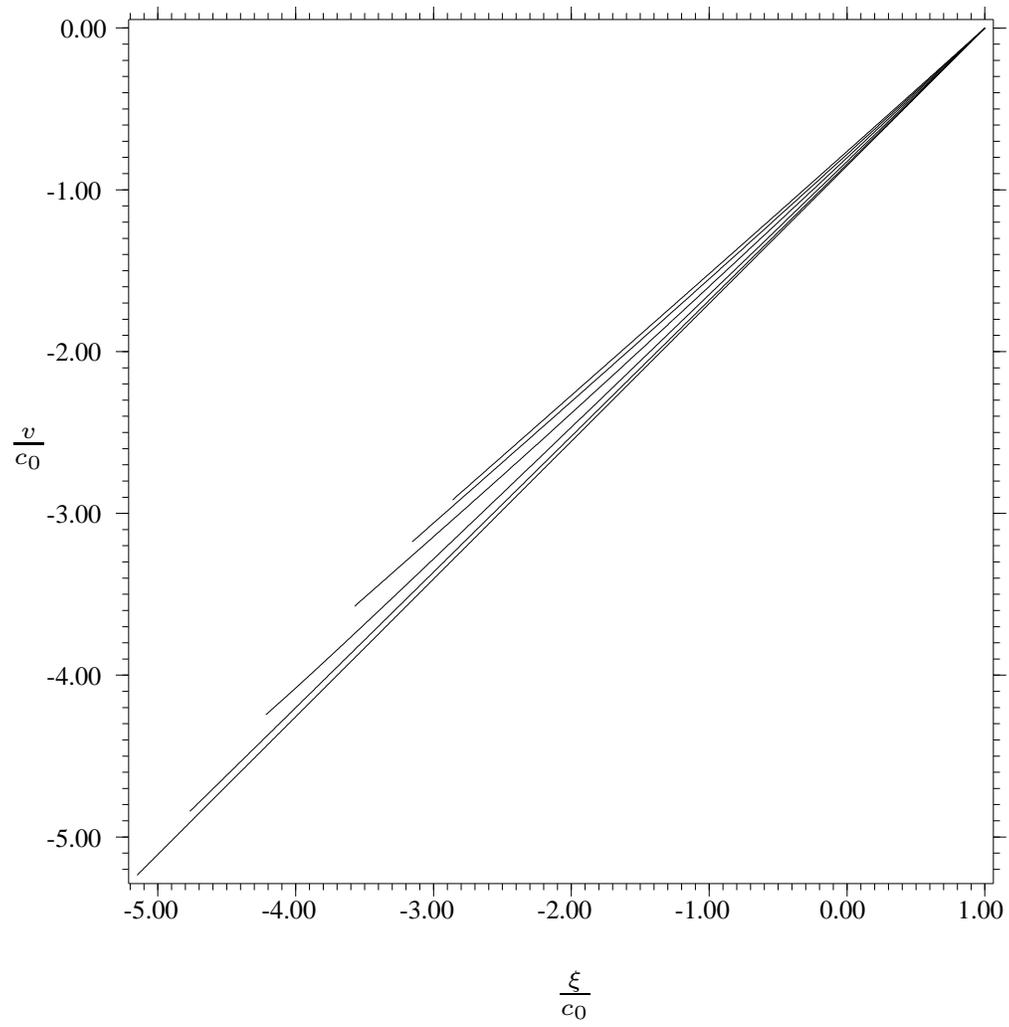}
\caption{
Velocity profile for $\gamma = 5/3$ and $\zo = 0.03$, $0.1$, $0.3$, $1$, $3$ and $10$ (from top to bottom). Results outside of this range are close to the bracketing results. 
}
\label{vprofile}
\end{figure}

\begin{figure}
\psfrag{y}[][][1.5]{$\frac{\rho}{\rho_0}$}
\psfrag{x}[][][1.5]{$\frac{\xi}{\cto}$}
\plotone{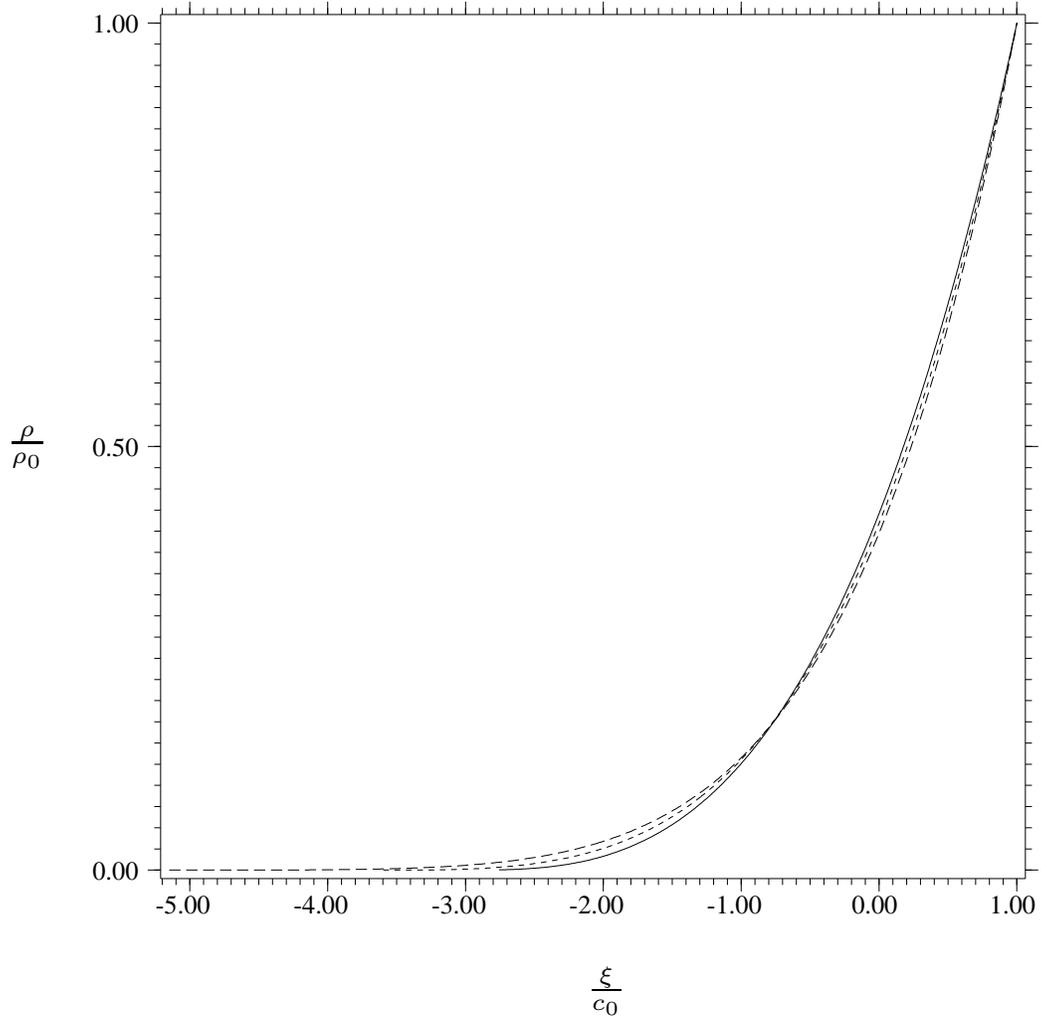}
\caption{
Density profile for $\gamma = 5/3$ and $\zo = 0.01$ ({\it solid line}), $0.3$ ({\it dotted line}) and $10$ ({\it dashed line}). Results outside of this range are close to the bracketing results. 
}
\label{rhoprofile}
\end{figure}

\begin{figure}
\psfrag{y}[][][1.5]{$\frac{T}{T_0}$}
\psfrag{x}[][][1.5]{$\frac{\xi}{\cto}$}
\plotone{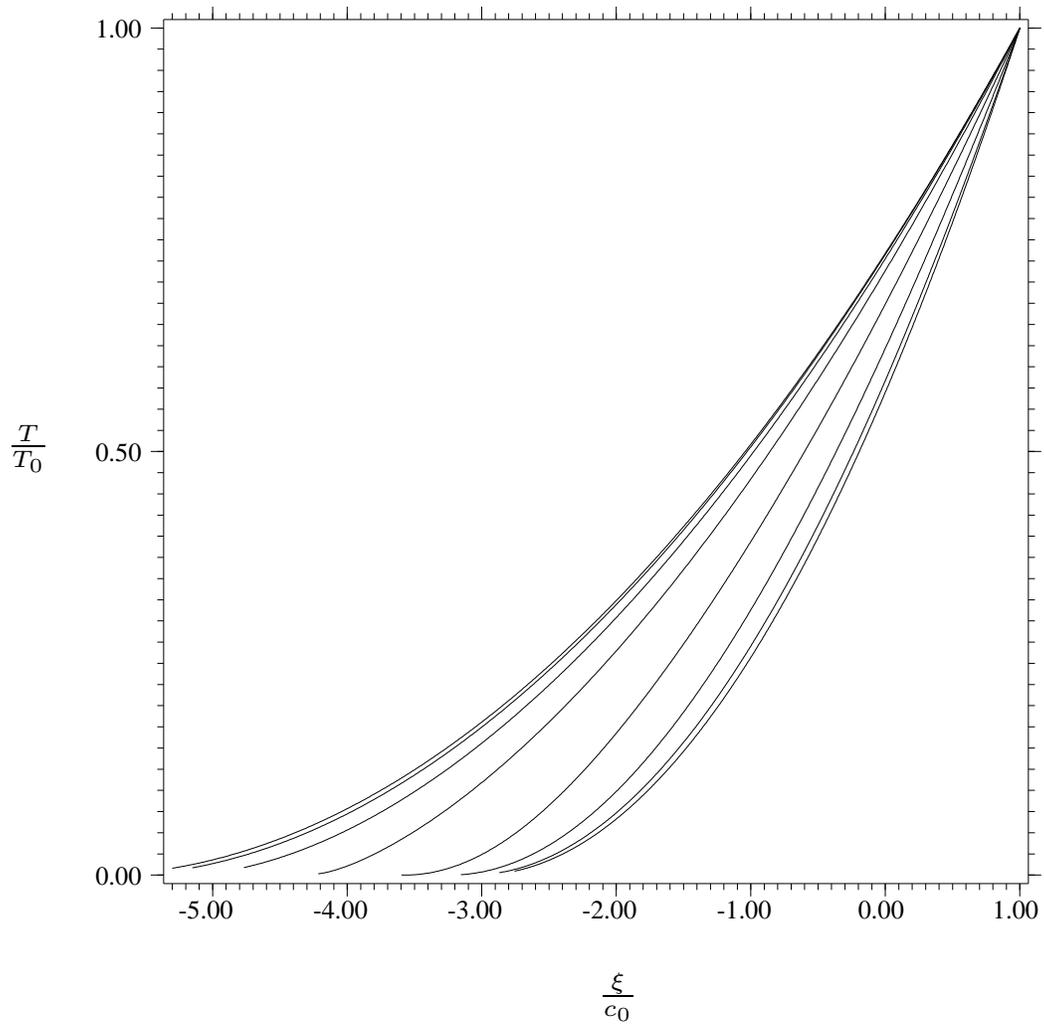}
\caption{
Temperature profile for $\gamma = 5/3$ and $\zo = 0.01$, $0.03$, $0.1$, $0.3$, $1$, $3$, $10$ and $30$ (from bottom to top). Results outside of this range are close to the bracketing results. 
}
\label{Tprofile}
\end{figure}

\begin{figure}
\psfrag{y}[][][1.5]{$\frac{\z}{\zo}$}
\psfrag{x}[][][1.5]{$\frac{\xi}{\cto}$}
\plotone{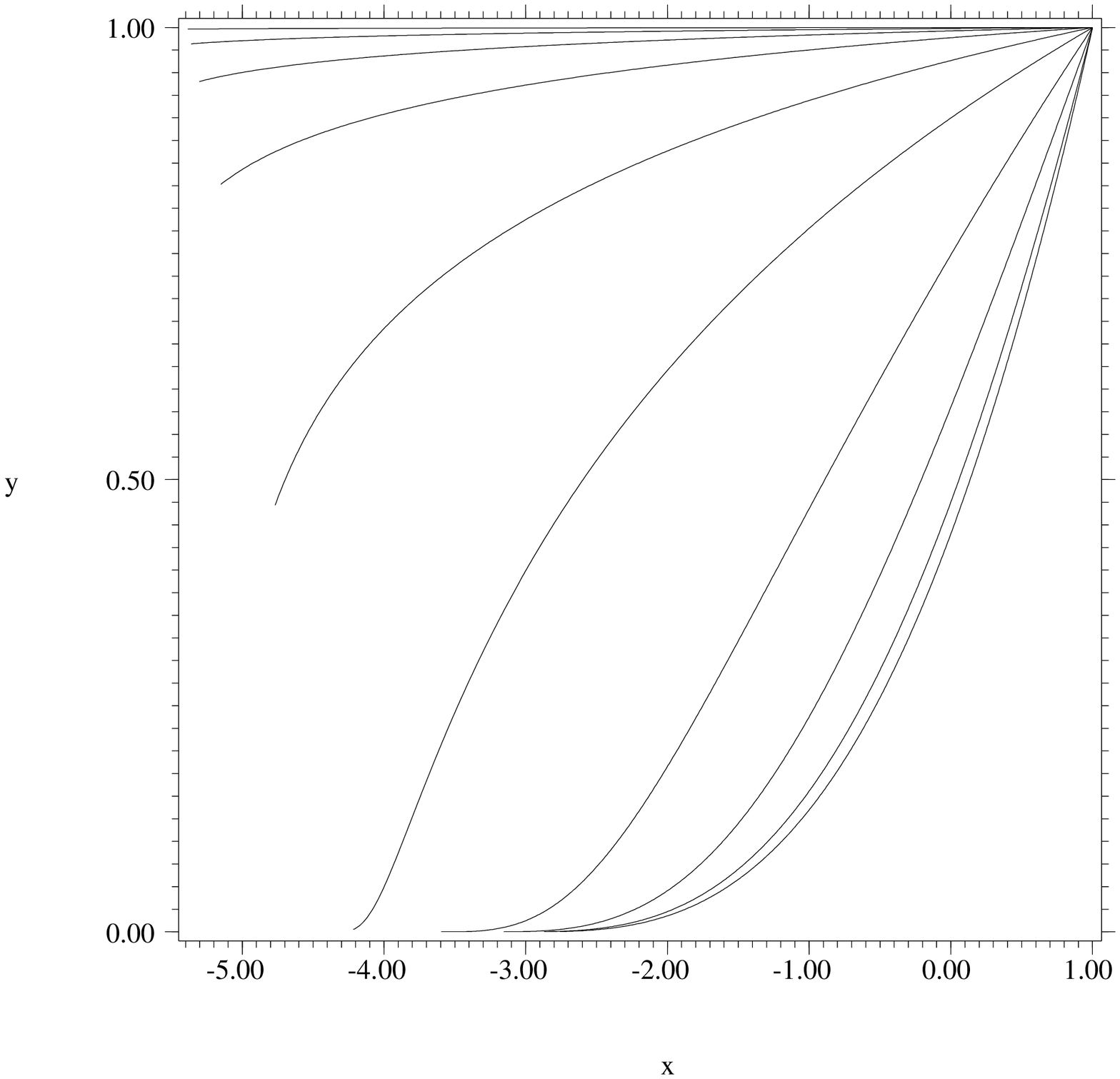}
\caption{
Profile of $\z$ for $\gamma = 5/3$ and $\zo = 0.01$, $0.03$, $0.1$, $0.3$, $1$, $3$, $10$, $30$, $100$ and $1000$ (from bottom to top). Results outside of this range are close to the bracketing results. 
}
\label{Zprofile}
\end{figure}

Analytical profiles of velocity, density, temperature and $\alpha$ as a function of $\xi$ for $\gamma = 5/3$ are shown in Figures~\ref{vprofile}-\ref{Zprofile}. The fact that $\z \sim const.$ for $\zo \gg 1$ is consistent with expression (\ref{TRHO}), which gives $\rho \propto T^3$ for $\z \gg 1$, i.e., $\z \propto T^3/\rho \sim const$; it is simply the conservation of entropy in the high energy density limit. When $\gamma = 4/3$, $\Gamma_1 = 4/3$, $\ct = \cs\sqrt{1 + \zo}$ and expressions (\ref{VAPPROX}) and (\ref{CAPPROX}) are exact.

\subsection{Isothermal Limit}

\begin{figure}
\psfrag{y}[][][1.5]{$\frac{\ct}{\cto}$}
\psfrag{x}[][][1.5]{$\frac{\xi}{\cto}$}
\plotone{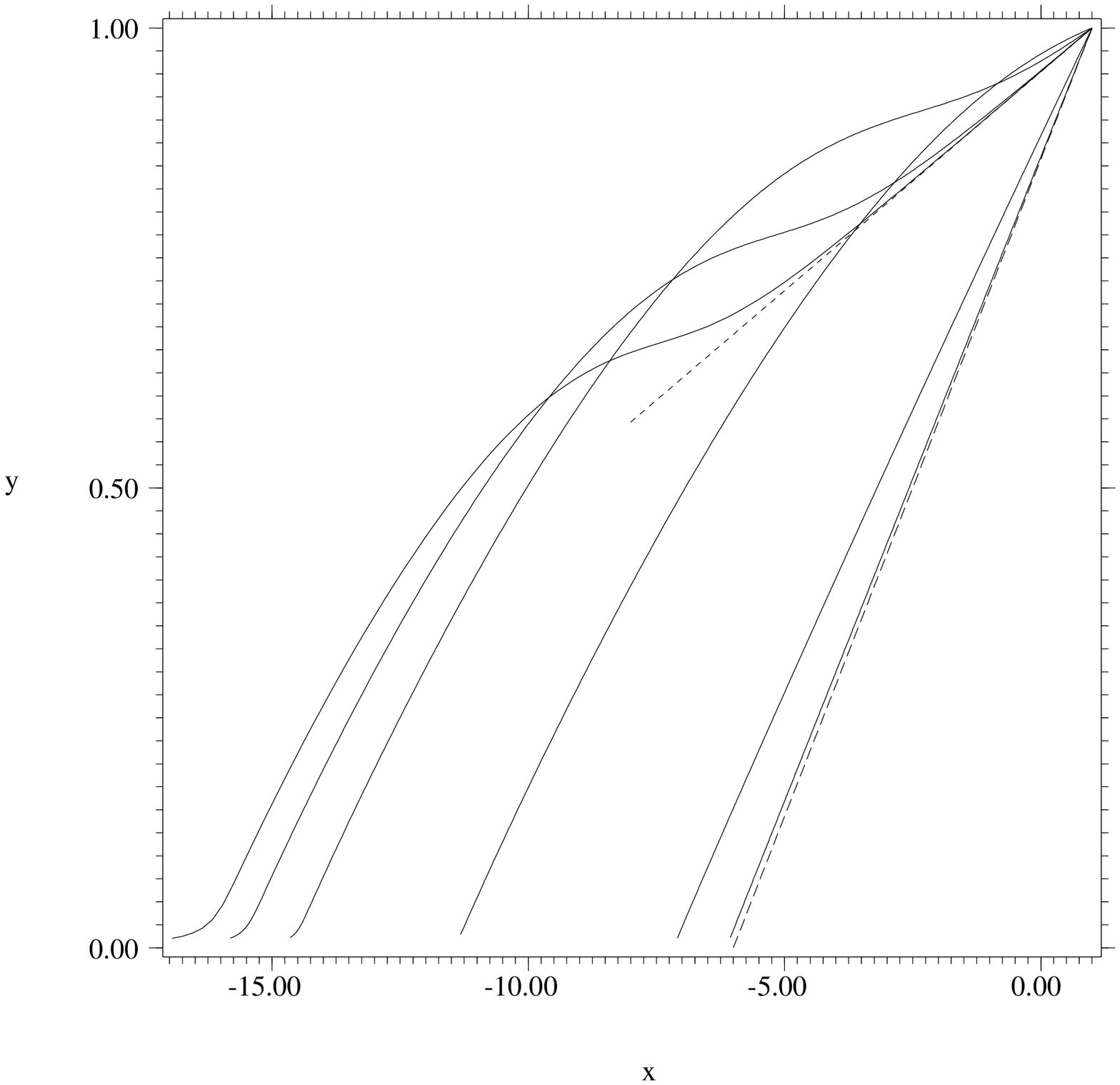}
\caption{
Profile of $\ct$ for $\gamma = 1.1$ and $\zo = 0.001$, $0.01$, $0.1$, $1$, $10$ and $100$ ({\it solid lines} from left to right). Also shown are the hydrodynamic results for $\gamma = 1.01$ ({\it dotted line}) and $\gamma = 4/3$ ({\it dashed line}). 
}
\label{ca1.1}
\end{figure}

\begin{figure}
\psfrag{y}[][][1.5]{$\frac{\ct}{\cto}$}
\psfrag{x}[][][1.5]{$\frac{\xi}{\cto}$}
\plotone{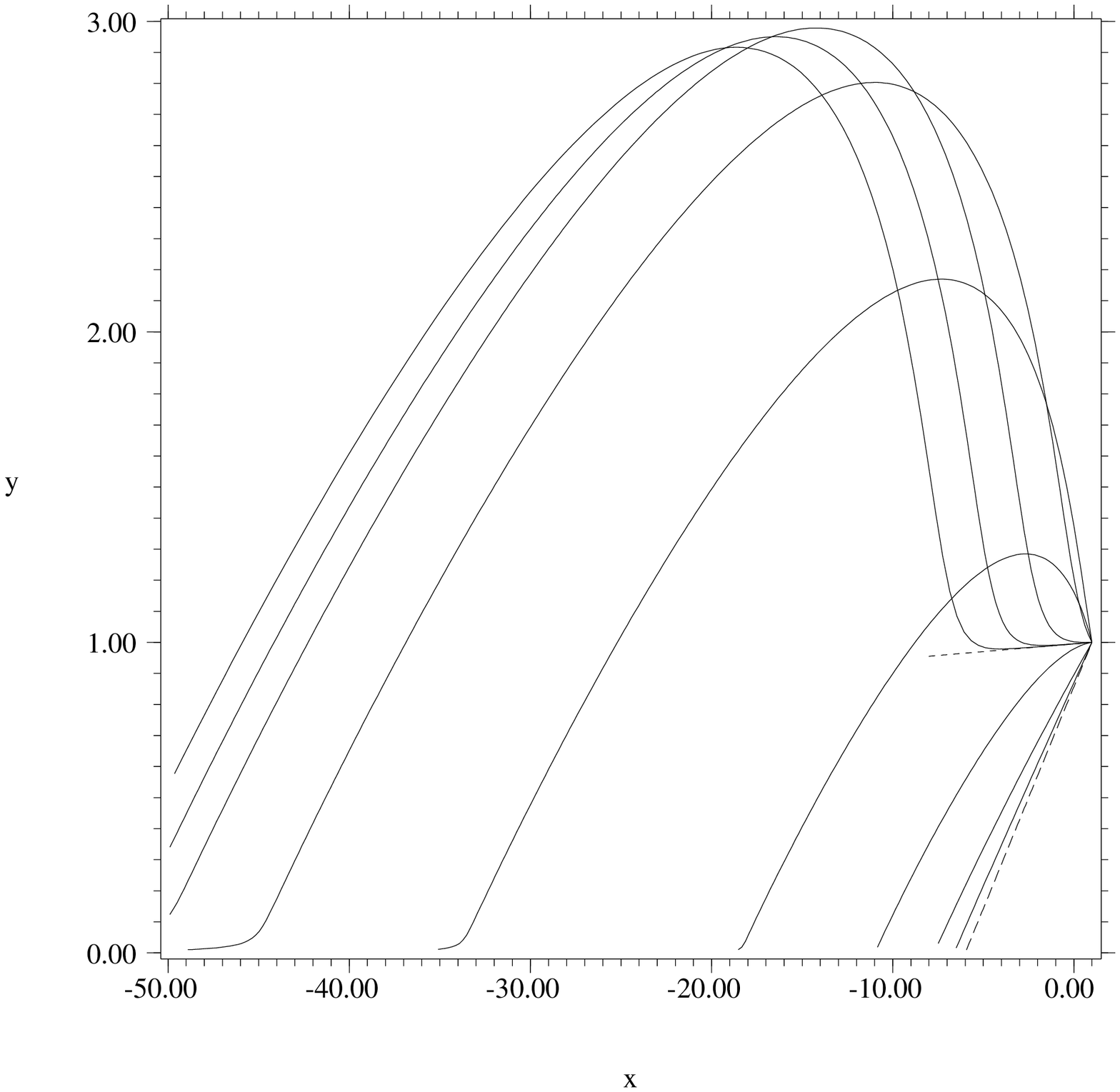}
\caption{
Profile of $\ct$ for $\gamma = 1.01$ and $\zo = 0.001$, $0.01$, $0.1$, $1$, $3$, $10$, $30$, $100$ and $300$ ({\it solid lines} from left to right). Also shown are the hydrodynamic results for $\gamma = 1.01$ ({\it dotted line}) and $\gamma = 4/3$ ({\it dashed line}). 
}
\label{ca1.01}
\end{figure}

The considerations that lead to the approximate expressions (\ref{VAPPROX}) and (\ref{CAPPROX}) break down when $\Gamma_1 \simeq 1$. In that case, and for a precise code comparison, the exact solution must be used (see the Appendix for details on its calculation). The fluid quantity that is most sensitive to the breakdown of the approximate solution is $\ct$. Figures~\ref{ca1.1} and \ref{ca1.01} show profiles of the adiabatic sound speed for $\gamma = 1.1$ and $\gamma = 1.01$. As the adiabatic index of the material approaches unity, the speed of sound in the fluid {\it increases} as it rarefies. This is generally associated with a value for $\Gamma_1$ less than unity,\footnote{This demonstrates another breakdown in the analogy between a radiating fluid and a material fluid with $\gamma = \Gamma_1$. If the latter were strictly true, $\Gamma_1 < 1$ would imply a negative pressure; as it is, despite the somewhat unusual behavior of the sound speed, the pressure remains perfectly well behaved.}  which from expression (\ref{GDEF}) occurs for $\z$ less than
\be
\z_{crit} = \frac{1}{8(\gamma - 1)}\left(1 + \sqrt{1 - \left[\frac{4(\gamma-1)}{9-8\gamma}\right]^2}\right) \simeq \frac{0.25}{\gamma - 1},
\ee
where the latter expression is valid for $\gamma - 1 \ll 1$. This is equivalent to a ratio of energy densities near unity. Real solutions to the above expression can only be obtained for $\gamma < 13/12$. The increase of $\ct$ in regions of rarefaction thus occurs for a material component with
\be
f = \frac{2}{\gamma - 1} > 24,
\ee
where $f$ is the microscopic degrees of freedom. A more careful analysis (see the Appendix) confirms the $f > 24$ result, with a slight modification to the value for $\z_{crit}$ (expression [\ref{ZCRIT}]). The calculation in the Appendix also demonstrates that 1) the peak in $\ct$ seen in Figure~\ref{ca1.01} continues to grow with decreasing $\gamma$, with $c_{a,max} \sim (\gamma - 1)^{-1/2}$ (expression [\ref{CMAX}]), 2) the sound speed always decreases for $\zo > \z_{crit}$ and 3) the portion of the solution that approaches the isothermal limit only exists for $\zo < 0.714$.

\begin{figure}
\psfrag{Y}[][][1.5]{$\z$}
\psfrag{X}[][][1.5]{$\frac{\xi}{\cto}$}
\plotone{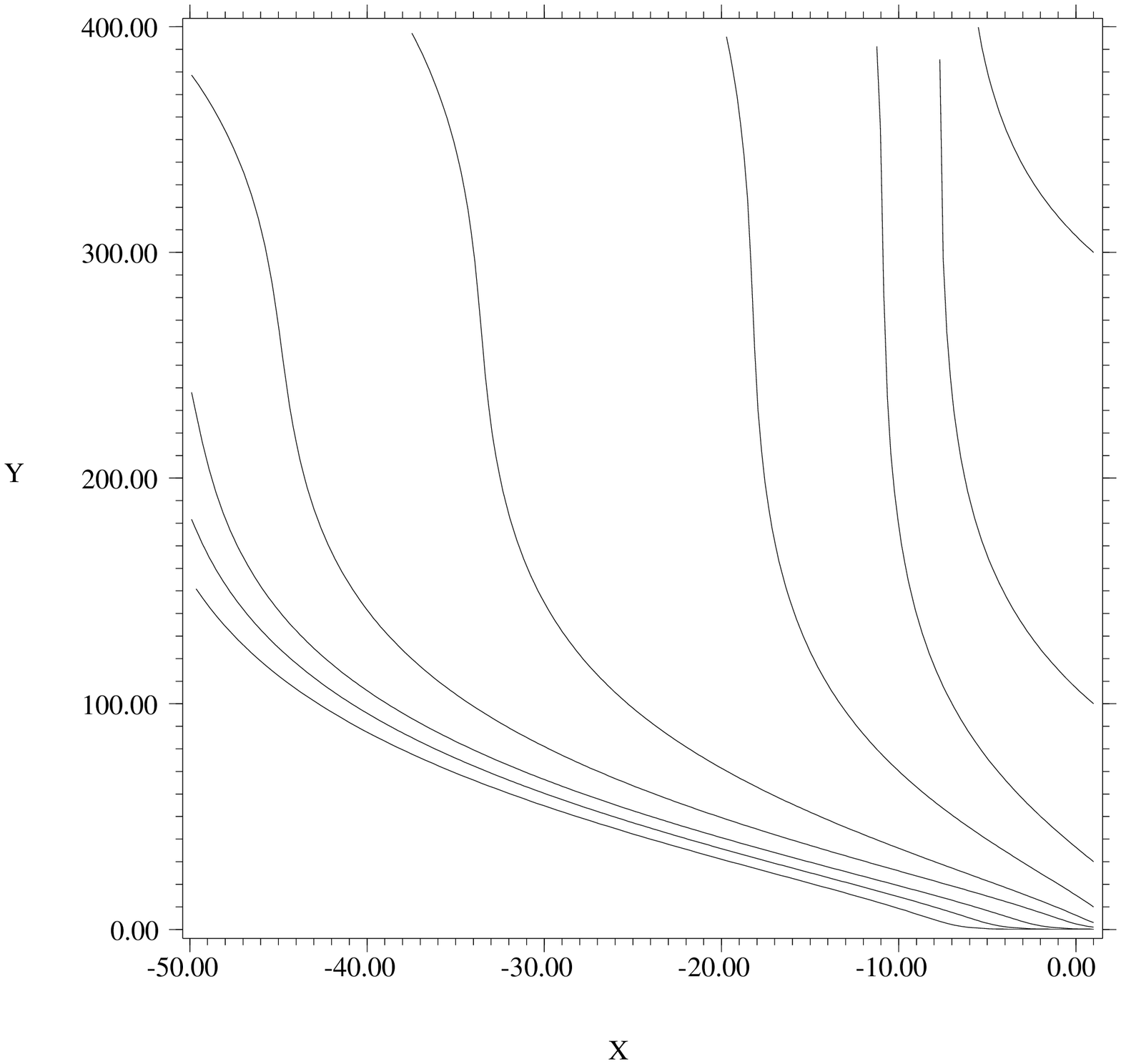}
\caption{
Profile of $P_r/P_g$ for $\gamma = 1.01$ and $\zo = 0.001$, $0.01$, $0.1$, $1$, $3$, $10$, $30$, $100$ and $300$ (from bottom to top).
}
\label{Z1.01}
\end{figure}

\begin{figure}
\psfrag{y}[][][1.5]{$\frac{E}{U}$}
\psfrag{x}[][][1.5]{$\frac{\xi}{\cto}$}
\plotone{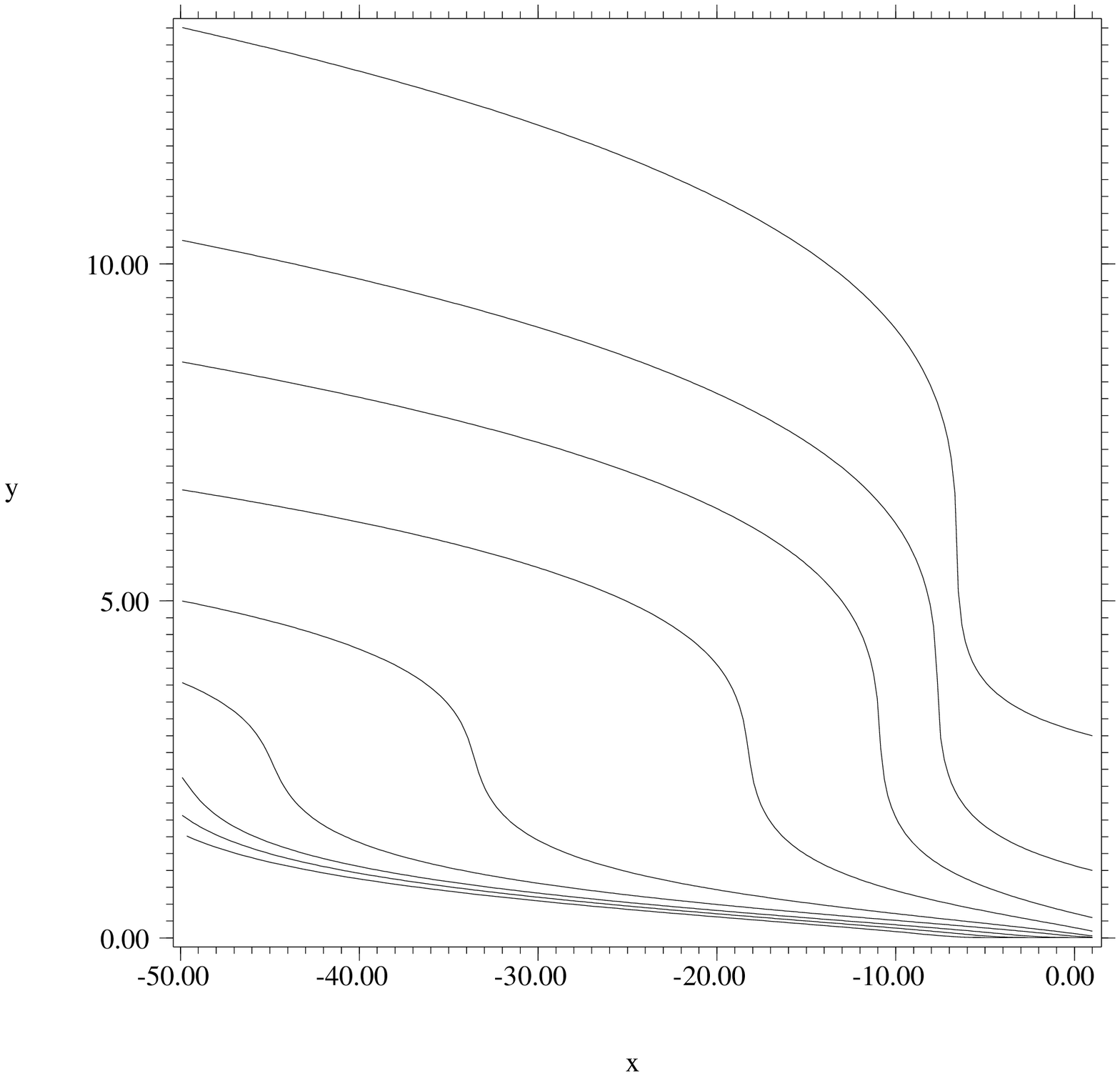}
\caption{
Profile of $E/U$ for $\gamma = 1.01$ and $\zo = 0.001$, $0.01$, $0.1$, $1$, $3$, $10$, $30$, $100$ and $300$ (from bottom to top).
}
\label{r1.01}
\end{figure}

Part of the reason for the unusual behavior of the sound speed is the increase of $\z$ as the fluid is rarefied (recall that this occurs for any $\gamma < 4/3$). Figure~\ref{Z1.01} shows the profile of $\z$ for $\gamma = 1.01$; even for $\zo \ll 1$, the pressure ratio increases rapidly in the rarefaction region and the fluid quickly becomes dominated by radiation pressure, independent of the initial pressure ratio. Figure~\ref{r1.01} shows the ratio of energy densities for the same set of parameters; this remains close to unity for a wider range of parameters. While the fluid near the quiescent state asymptotes to the hydrodynamic solution ($\Gamma_1 \simeq 1$), the fluid eventually (for a sufficiently large piston velocity) approaches the radiation-dominated solution ($\Gamma_1 = 4/3$) in which {\it both} radiation pressure and radiation energy density dominate. The increase of $\ct$ occurs in the transition between these two asymptotic regimes where the fluid is dominated by radiation pressure but the material and radiation energy densities are comparable. One clear implication of these results is that RHD calculations with $\gamma \simeq 1$ can produce drastically different results from a strictly isothermal calculation.

\subsection{Effects of Diffusion}

\begin{figure}
\psfrag{y}[][][1.5]{$\frac{v}{\cto}$}
\psfrag{x}[][][1.5]{$\frac{\xi}{\cto}$}
\plotone{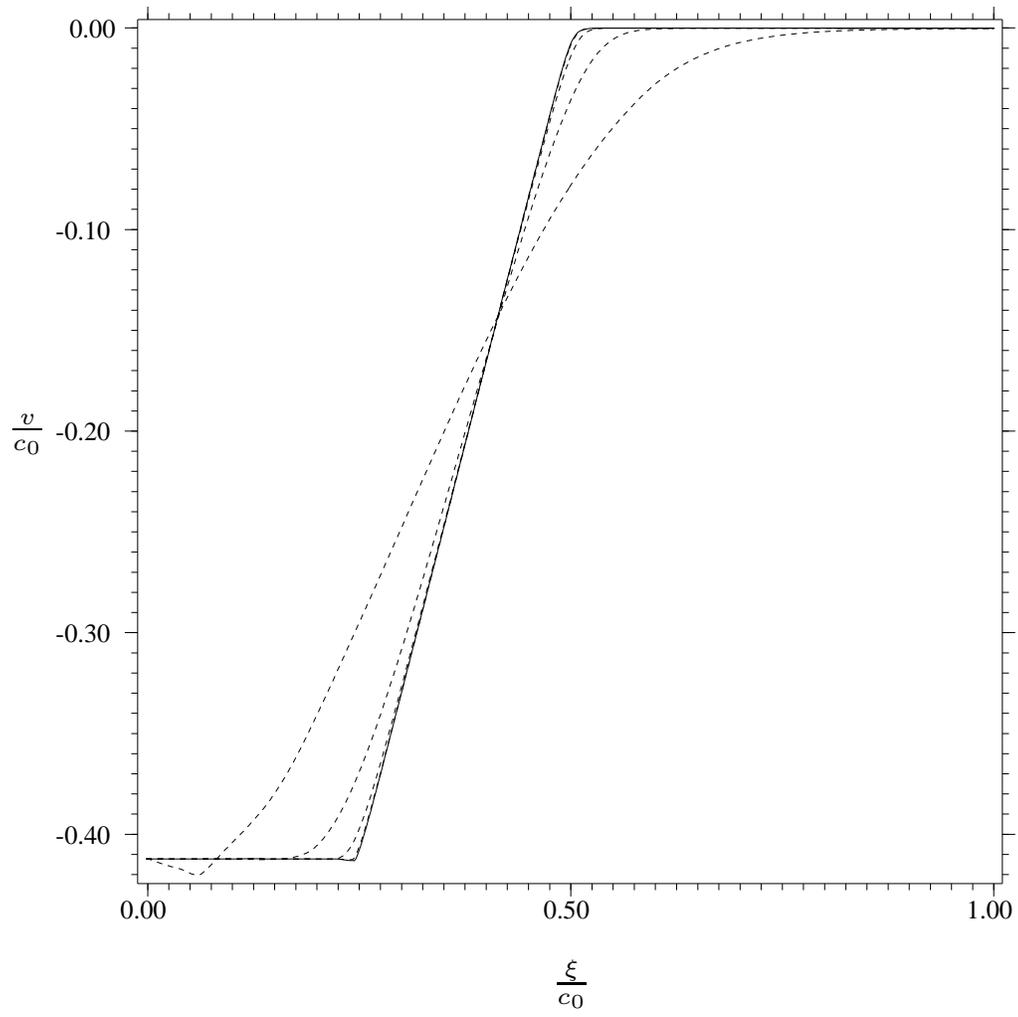}
\caption{
Velocity profile from a finite difference numerical integration with $\gamma = 1.6667$, $\zo = 1$ and $\beta_0 \tau_0 = \infty$ ({\it solid line}), $10^4$, $10^3$, $10^2$ and $10$ ({\it dashed lines} in order of decreasing accuracy). For $\beta_0 \tau_0 = 10$, the temperature drop is sufficient to make $\beta \tau \sim 1$ at $x \sim 0.1L$.
}
\label{diffusion}
\end{figure}

Figure~\ref{diffusion} shows the velocity profile from finite difference numerical integrations both with and without diffusion, for various values of $\beta_0 \tau_0$. The parameters for these calculations are $\gamma = 1.6667$, $\beta_0 = 10^{-4}$ and $\zo = 1$. For sufficiently large values of $\beta_0 \tau_0$, the diffusion is only modifying the transition between the self-similar and constant portions of the fluid, as expected. 

\section{Characteristics and Reimann Invariants}\label{CRI}

As discussed in \S101 of \cite{ll87}, all that is required to extend the results of the previous section to more general boundary conditions is the ability to express all of the fluid quantities as functions of one another ($P = P[\rho]$, $v = v[P]$, etc.). This is ensured for isentropic boundary conditions, and the generalization of expression (\ref{CHAR}) is
\be
x = t\left(v \pm \ct\right) + f(v),
\ee
where $f(v)$ is an arbitrary function of the velocity. These solutions are referred to as simple waves, since the flow variables are all functions of $x \pm \ct t$. The centered rarefaction wave is a simple wave with $f(v)= 0$.

The Riemann invariants for the combined fluid are
\be
J_\pm = v \pm \int \frac{dP}{\rho \ct}.
\ee
They are conserved along the characteristic curves $C_\pm = v \pm \ct$. As discussed in \S104 of \cite{ll87}, they are strictly conserved only for isentropic flow. For adiabatic flow, the perturbations $\delta v \pm \delta P/(\rho \ct)$ are conserved along the $C_\pm$ characteristics, and perturbations in the total entropy, $\delta s_m + \delta s_r \propto (\delta U + \delta E - [\gamma U + 4E/3] \delta \rho/\rho)/U \propto (1/[\gamma - 1] + 12\z)\delta T/T - (1+ 4\z) \delta \rho/\rho$, are conserved along the characteristic $C_0 = v$. 

\section{Summary and Discussion}\label{SD}

This paper has regarded a radiating fluid in the dynamic diffusion limit as ideal; diffusion is neglected for the bulk of the fluid since it only modifies regions of the flow in which gradients are large, in a manner analogous to other forms of dissipation such as viscosity and heat conduction. A complete analysis has been performed for a centered rarefaction wave (\S\ref{CRW}), and expressions for the characteristic curves and Reimann invariants of ideal RHD have been provided (\S\ref{CRI}). It has been shown that the much of the hydrodynamic analysis carries over by simply replacing the standard adiabatic quantities with the form that they take in a radiating fluid (\S\ref{THERMO}). The exact solution requires the numerical integration of expression (\ref{V}) and the implicit solution of equation (\ref{CHAR}).
The qualitative nature of the solution changes as the internal degrees of freedom of the material become significant ($f > 24$). In that case, if the initial energy content of the radiation is less the $90\%$ of the internal energy of the material, the sound speed of the fluid increases as it rarefies.

One application of \S\ref{CRW} is as a test of numerical RHD codes; the solutions are nonlinear and the only term that has been neglected is the diffusion operator. One can ensure that the system of equations is in the dynamic diffusion limit simply by setting the length of the computational domain such that its optical depth satisfies $\tau \gg c/\ct$. Other tests could be constructed by further extending hydrodynamic results; to cite a couple of examples, \cite{ll87} include analyses of a uniformly accelerated piston and a centered rarefaction wave reflecting off a solid wall. The solution for a compressive wave before it steepens into a shock can be obtained from the negative branch of equations (\ref{CHAR}) and (\ref{V}).

Any application of the results of \S\ref{CRW} to physical systems must keep in mind the idealized nature of the analysis. Since the density and temperature of the solutions change significantly from their quiescent values, there are likely to be large regions of parameter space for which the assumptions of dynamic diffusion break down. Mapping out the limits of their validity for realistic opacities would be a useful follow-on exercise to this work. For flows with $\beta \tau \sim 1$, the modification of these solutions by diffusion in the transition regions could likely be obtained by an asymptotic analysis.

The considerations of \S\ref{CRI} could be used in the development of Godunov numerical schemes for RHD. An analysis of the hyperbolic nature of the full set of RHD equations has been conducted by \cite{bal99a,bal99b}; the radiation flux in these papers was regarded as a source term, however, so that only the material sound speed enters the expression for the $C_\pm$ characteristics. The analysis performed here points to a more self-consistent approach for numerically solving the equations of RHD in the dynamic diffusion limit. For calculations in which both advection and diffusion are important, one could still employ a Reimann solver for the hyperbolic portion of the equations while treating the diffusion separately \citep{dnd06}.

\acknowledgements

I am grateful to Dimitri Mihalas, John Castor, Richard Klein and Charles Gammie for their comments. This work was performed under the auspices of Lawrence Livermore National Security, LLC, (LLNS) under Contract No.$\;$DE-AC52-07NA27344.

\begin{appendix}

\section{Calculation Details}

The velocity can be calculated as a function of any of the other fluid variables by expressing the integral (\ref{V}) in the form appropriate for that variable. The closed form expression of the integral in terms of $\z$ is given by 
\be
v = \int \ct \, \frac{1 + 12(\gamma - 1)\z}{(3\gamma - 4)\z} \, d\z,
\ee
where $\ct$ is given by expression (\ref{CT}). Numerical evaluation of this integral is straightforward for $\gamma > 4/3$, but it becomes somewhat problematic when $\gamma < 4/3$ due to the rapid increase of $\z$.\footnote{To get the rarefaction solution, one must integrate between $\z = \zo$ and $\z \lessgtr \zo$ when $\gamma \gtrless 4/3$.} Expressing the integral in terms of the logarithm of the density appears to be more robust, although other choices may be superior. Equation (\ref{RHOZ}) can be inverted to express $\z$ in terms of the density:
\be\label{ZRHO}
A\frac{\z}{\zo} = W\left(A \exp\left[A + (3\gamma - 4) \ln \frac{\rho}{\rho_0}\right]\right),
\ee
where $W$ is the Lambert-$W$ function (or product log) and
\be
A \equiv 12(\gamma - 1)\zo.
\ee
The velocity is then given by
\be
v = \int \ct\left(\eta\right) \, d\eta,
\ee
where $\eta = \ln(\rho/\rho_0)$ and $\ct(\eta)$ is obtained from expressions (\ref{CT}) and (\ref{ZRHO}).
The flow profiles in the text were plotted with Mathematica, which provides a module for evaluating the product log, although one can generate its values based upon a simple recursion formula. With $v(\rho)$ obtained numerically, equation (\ref{CHAR}) was solved with a root-finding algorithm to obtain $\rho(\xi)$. All the fluid quantities as a function of $\xi$ then follow from the expressions given in the text.

The unusual behavior of the sound speed as $\gamma \rightarrow 1$ discussed in the text can be obtained analytically as follows. Clearly an increase in the sound speed requires that there be one or more locations in the fluid for which $d\ct/d\xi = 0$. Differentiating expression (\ref{CT}), one finds that
\be
\frac{1}{\ct}\dv{\ct}{\z} = 2\frac{(\gamma - 1)(1 + 4\z)}{1+A}\left(\frac{1+A}{4(3\gamma - 4)\z}
+ \frac{5 - 3\gamma + A}{1 + A + (\gamma - 1)(1 + 4\z)^2}\right).
\ee
Setting this expression to zero gives an expression that is quadratic in $\gamma - 1$ and cubic in $\zo$:
\be\label{EA}
(1+12\epsilon\z) (1 + 12\epsilon\z + \epsilon(1 + 4\z)^2)
+ 4(3\epsilon - 1)\z (2 - 3\epsilon + 12\epsilon\z) = 0,
\ee
where
\be
\epsilon \equiv \gamma - 1.
\ee
The physical branch of expression (\ref{EA}) regarded as an equation for $\epsilon(\z)$ is
\be
\epsilon = \frac{-1 - 68\z + 32\z^2 + (1 + 4\z)\sqrt{1 + 224\z + 448\z^2}}{48\z(-1 + 16\z + 8\z^2)}.
\ee
The maximum of the above expression occurs at $\z = 0.714$ and is $0.0862$; i.e., $\gamma < 1.0862$. This corresponds to $f > 23.2$, or, restricting $f$ to integer values, $f > 24$.

Solving expression (\ref{EA}) for $\z(\epsilon)$ yields two positive real roots when $\epsilon < 0.0862$. One of these corresponds to the peak value of $\ct$ and is approximately
\be\label{ZCRIT}
\z_{crit} \simeq \frac{0.304}{\epsilon} - 2.22 + 1.15\epsilon + O(\epsilon^2).
\ee
This corresponds to a ratio of energy densities
\be
\frac{E}{U} = 3\epsilon \z \simeq 0.9.
\ee
An estimate of the peak value of $\ct$ can be obtained by looking at its asymptotic form for $\epsilon \ll 1$ and $\z \epsilon \sim 1$:
\be
\ct \simeq 4\z\sqrt{\frac{\epsilon}{1+12\epsilon\z}} e^{-2\z\epsilon}.
\ee
This has a maximum at $\z\epsilon = (1+\sqrt{7})/12 = 0.304$ given by
\be\label{CMAX}
c_{a,max} \simeq \frac{0.307}{\sqrt{\epsilon}}.
\ee

The other root varies between $0.125 < \z < 0.714$ for $0 < \epsilon < 0.0862$. This corresponds to a minimum in $\ct$ and marks the transition between the portion of the solution that approaches the isothermal limit and the portion in which the sound speed increases. For $\zo > \z_{crit}$, neither of the roots is accessible (since $\z > \zo$). For $0.714 < \zo < \z_{crit}$, the first root is accessible but not the second. For $\zo < 0.714$ both roots are accessible.

\end{appendix}

\newpage

\begin{thebibliography}

\bibitem[Balsara(1999a)]{bal99a} Balsara, D.~S.\ 1999, Journal of Quantitative Spectroscopy and Radiative Transfer, 61, 629 

\bibitem[Balsara(1999b)]{bal99b} Balsara, D.~S.\ 1999, Journal of Quantitative Spectroscopy and Radiative Transfer, 61, 617 

\bibitem[Bell \& Lin(1994)]{bl94} Bell, K.~R., \& Lin, D.~N.~C.\ 1994, \apj, 427, 987 

\bibitem[Chandrasekhar(1967)]{ch67} Chandrasekhar, S.\ 1967, New York: Dover

\bibitem[Cox \& Giuli(1968)]{cg68} Cox, J.~P., \& Giuli, R.~T.\ 1968, New York: Gordon and Breach

\bibitem[Dragojlovic et al.(2006)]{dnd06} Dragojlovic, Z., Najmabadi, F., \& Day, M.\ 2006, Journal of Computational Physics, 216, 37 

\bibitem[Landau \& Lifshitz(1987)]{ll87} Landau, L.~D., \& Lifshitz, E.~M.\ 1987, Fluid Mechanics (2nd ed.), Oxford: Butterworth-Heinemann

\bibitem[Mihalas \& Mihalas(1984)]{mm84} Mihalas, D., \& Weibel Mihalas, B.\ 1984, New York: Oxford University Press, 1984

\bibitem[Stone \& Norman(1992)]{sn92} Stone, J.~M.~\& Norman, M.~L.\ 1992, \apjs, 80, 753

\bibitem[Stone et al.(1992)]{smn92} Stone, J.~M., Mihalas, D., \& Norman, M.~L.\ 1992, \apjs, 80, 819 

\bibitem[Turner \& Stone(2001)]{ts01} Turner, N.~J., \& Stone, J.~M.\ 2001, \apjs, 135, 95 

\end{thebibliography}
\end{document}